\documentclass[10pt,letterpaper]{article}
\usepackage{opex3}
\usepackage{epstopdf}
\usepackage{amsmath,bm}  
\usepackage{cite} 
\usepackage{diagbox}

\usepackage{graphicx}
\usepackage{subfigure}
\usepackage{fancyhdr}
\usepackage{xcolor}

\usepackage{yhmath}

\begin{document}

\title{First-principle calculation of Chern number in gyrotropic photonic crystals}

\author{Ran Zhao,$^{1,2}$ Guo-Da Xie,$^{1,2}$ Menglin L. N. Chen,$^{3}$ Zhihao Lan,$^{4}$ Zhixiang Huang,$^{1,2}$ and Wei E. I. Sha$^{5,*}$}

\address{$^1$ Key Laboratory of Intelligent Computing and Signal Processing, Ministry of Education, Anhui University, Hefei 230039, China\\
$^2$ Key Laboratory of Electromagnetic Environmental Sensing, Department of Education of Anhui Province, Hefei 230039, China\\
$^3$ Department of Electrical and Electronic Engineering, The University of Hong Kong, Hong Kong\\
$^4$ Department of Electronic and Electrical Engineering, University College London, United Kingdom \\
$^5$ Key Laboratory of Micro-nano Electronic Devices and Smart Systems of Zhejiang Province, College of Information Science Electronic Engineering, Zhejiang University,
 \\Hangzhou 310027, China\\

}

\quad\quad \quad\quad\quad\quad\quad\quad\quad \textcolor{red}{Chern number calculation within 15 seconds}
\homepage{http://www.zjuisee.zju.edu.cn/weisha/Publications/Files/Chern\_number.zip}
\homepage{https://doi.org/10.1364/OE.380077}

\email{$^{*}$weisha@zju.edu.cn} 


\begin{abstract}
 As an important figure of merit for characterizing the quantized collective behaviors of the wavefunction, Chern number is the topological invariant of quantum Hall insulators. Chern number also identifies the topological properties of the photonic topological insulators (PTIs), thus it is of crucial importance in PTI design. In this paper, we develop a first principle computatioal method for the Chern number of 2D gyrotropic photonic crystals (PCs), starting from the Maxwell's equations. Firstly, we solve the Hermitian generalized eigenvalue equation reformulated from the Maxwell's equations by using the full-wave finite-difference frequency-domain (FDFD) method. Then the Chern number is obtained by calculating the integral of Berry curvature over the first Brillouin zone. Numerical examples of both transverse-electric (TE) and transverse-magnetic (TM) modes are demonstrated, where convergent Chern numbers can be obtained using rather coarse grids, thus validating the efficiency and accuracy of the proposed method.
\end{abstract}

\section{Introduction}
Topology studies the invariant properties of geometry under continuous deformation\cite{armstrong2013basic}. While in mathematics, topological invariants are commonly used to classify topological spaces, in topological physics, topological invariants are explored to distinguish the bulk properties of materials. If physical observables can be expressed as topological invariants, they can only vary discretely and will not be affected by small perturbations of system parameters.

With the discovery of the quantum Hall effects\cite{hasan2010colloquium,zhang2011topological} and recent advances in the study of topological insulators\cite{klitzing1986the}, topological phases of matter have attracted great attention in condensed matter physics.
In 2005, Haldane and Raghu transferred the key feature of quantum Hall effect in quantum mechanics to classical electromagnetics\cite{haldane2008possible} and soon after, it was numerically and experimentally verified by using photonic crystals (PCs)\cite{wang2008reflection,wang2009observation}.

Chern number in a photonic system is defined on the dispersion bands in wave-vector space. For a two-dimensional (2D) periodic system, the Chern number is the integration of the Berry curvature over the first Brillouin zone. Once the Chern number is calculated, the topological properties of the system can be identified (trivial or non-trivial). In addition, the Chern number can be used to explain the phenomenon of ``topological protection'' of the edge state transmission in photonic topological insulators (PTIs). Therefore, the accurate computation of the Chern number is of crucial importance in the PTI design\cite{lu2014topological,hassani2017berry,xie2018photonics}.

In this paper, we propose a first-principle computation method for Chern number calculation, based on the finite-difference frequency-domain (FDFD) method\cite{zhu2003full,guo2004photonic,Chen2018Generation,Fang2017Maxwell}. Compared with the commonly used plane-wave expansion (PWE) method\cite{Ho1990existence,Meade1993accurate,leung1990full}, the FDFD method is not only accurate and stable but also has lower computational complexity. Firstly, the FDFD method is used to compute the band structure of 2D gyrotropic PCs by solving the generalized eigenvalue equations derived from Maxwell's equations. Then, the formulae for the numerical calculation of the Chern number are derived in the discretized first Brillouin zone. At last, numerical examples are given to demonstrate the accuracy of the proposed method. Note, in this work, we will focus on 2D PCs with lossless, non-dispersive, local materials, which is a common practice in computing Chern numbers. The effects of material dispersion and loss in a real microwave ferrite has been discussed in \cite{wang2008reflection}. One remark we would like to make is that as the Chern number is a global property of the energy band, it does not depend on the local changes of the band caused by material dispersion as long as the band gap is still open, which also means that the Chern number has a certain degree of built-in robustness against material dispersion. Nevertheless, the extension of the current method to fully include the effect of material dispersion on the energy bands and Chern number calculations is an interesting direction for future research.

\section{Generalized eigenvalue problem using FD method}

\subsection{Maxwell's equations in the generalized coordinates}

\begin{figure}[htbp]
\centering
\includegraphics[width=8cm]{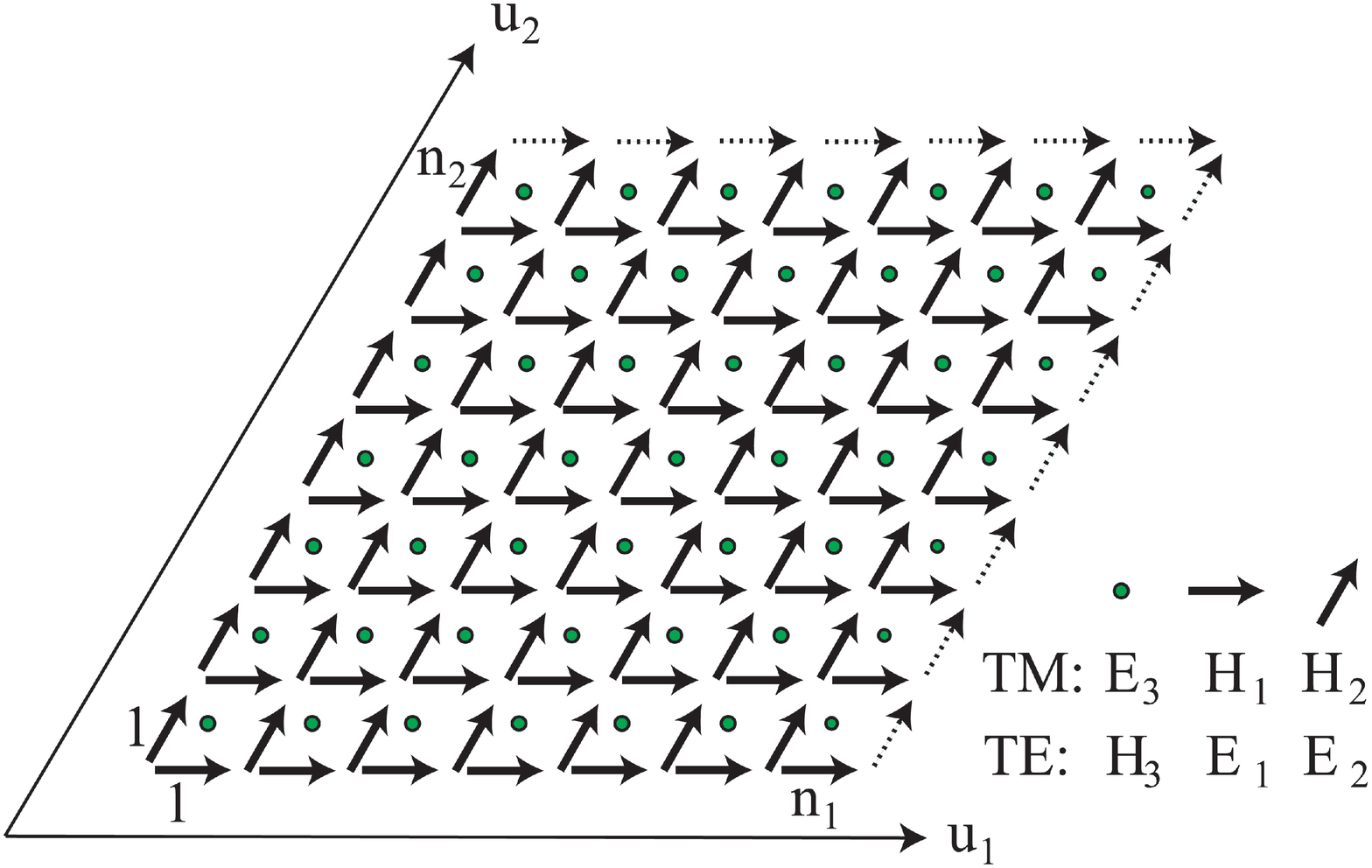}
\caption{The 2D Yee's grid in general coordinates. The dashed arrows are where the periodic boundaries apply.}
\label{FIG1}
\end{figure}

Firstly, the Maxwell's curl equations in free space in the generalized coordinate (Fig.~\ref{FIG1}) are expressed by

\begin{equation} \label{EQ1}
\nabla_{q}\times\hat{\mathbf{H}}=\imath k_0\mathbf{\hat{\epsilon}}(\mathbf{r})\hat{\mathbf{E}},\quad
\nabla_{q}\times\hat{\mathbf{E}}=-\imath k_0\mathbf{\hat{\mu}}(\mathbf{r})\hat{\mathbf{H}}.
\end{equation}
Here, $\nabla_{q}$ is the partial differential  operator in the generalized coordinate with three unit vectors $\hat{\bm u}_q(x,y,z),(q=1,2,3)$. $\hat{\mathbf{E}}$ and $\hat{\mathbf{H}}$ are the normalized vector fields with $\hat{E}_i=Q_i\sqrt{\frac{\epsilon_0}{\mu_0}}E_i, \quad\hat{H}_i=Q_iH_i(i=1,2,3)$. Actually, in the numerical implementation, $Q_i$s are the sizes of the discrete grids along each direction\cite{{Ward1996refraction}}. $k_0$ is the free-space wavenumber. $\hat{\epsilon}$ and $\hat{\mu}$ are the relative permittivity and permeability in the form of $3\times3$ tensor which are written as

\begin{equation} \label{EQ2}
\mathbf{\hat{\epsilon}}_{ij}(\mathbf{r})=[\bar{\bar{\epsilon}}(\mathbf{r})\cdot\mathbf{g}]_{ij}  \cdot|\hat{\bm u}_1\cdot \hat{\bm u}_2\times{\hat{\bm u}_3}|\frac{Q_1Q_2Q_3}{Q_iQ_j};\quad
\mathbf{\hat{\mu}}_{ij}(\mathbf{r})=[\bar{\bar{\mu}}(\mathbf{r})\cdot\mathbf{g}]_{ij}  \cdot|\hat{\bm u}_1\cdot \hat{\bm u}_2\times{\hat{\bm u}_3}|\frac{Q_1Q_2Q_3}{Q_iQ_j};
\end{equation}
where $\bar{\bar{\epsilon}}(\mathbf{r})$ and $\bar{\bar{\mu}}(\mathbf{r})$ are the original relative permittivity and permeability in the Cartesian basis. $\mathbf{g}$ is the metric tensor matrix which can be expressed as,

\begin{equation} \label{EQ3}
\mathbf{g}=
\left[
\begin{array}{cccc}
\hat{\boldsymbol{u}}_1\cdot \hat{\bm u}_1&\hat{\bm u}_1\cdot \hat{\bm u}_2&\hat{\bm u}_1\cdot \hat{\bm u}_3\\
\hat{\bm u}_2\cdot \hat{\bm u}_1&\hat{\bm u}_2\cdot \hat{\bm u}_2&\hat{\bm u}_2\cdot \hat{\bm u}_3\\
\hat{\bm u}_3\cdot \hat{\bm u}_1&\hat{\bm u}_3\cdot \hat{\bm u}_2&\hat{\bm u}_3\cdot \hat{\bm u}_3\\
\end{array}
\right]^{-1}.
\end{equation}

With the metric tensor matrix, the length of a vector in the generalized coordinate is calculated by

\begin{equation} \label{EQ4}
\left| \mathbf{r} \right|^2= \mathbf{r}^\text{T} \cdot\mathbf{g}^{-1}\cdot \mathbf{r}.
\end{equation}

\subsection{Bloch's theorem }

In 2D PCs, due to the translational symmetry, the modes can be written in the Bloch form:

\begin{equation} \label{EQ5}
\begin{array}{c}
\mathbf{E}_{\left(n, k_{z}, \mathbf{k}_{ \|}\right)}(\mathbf{r})=e^{-i \mathbf{k}_{ \|} \cdot \boldsymbol{\rho}} e^{-i k_{z} z} \mathbf{u}_{e \left(n, k_{z}, \mathbf{k}_{ \|}\right)}(\boldsymbol{\rho}),\\
\mathbf{H}_{\left(n, k_{z}, \mathbf{k}_{ \|}\right)}(\mathbf{r})=e^{-i \mathbf{k}_{ \|} \cdot \boldsymbol{\rho}} e^{-i k_{z} z} \mathbf{u}_{h \left(n, k_{z}, \mathbf{k}_{ \|}\right)}(\boldsymbol{\rho}).
\end{array}
\end{equation}
Here, $n$ is the band number, $\boldsymbol{\rho}$ is the mapping of vector $\mathbf{r}$ in the $(u_1, u_2)$ plane. $\mathbf{u}_e( \boldsymbol{\rho})$ and $\mathbf{u}_h(\boldsymbol{\rho})$ are periodic functions satisfying $\mathbf{u}_h(\boldsymbol{\rho})=\mathbf{u}_h(\boldsymbol{\rho}+\mathbf{R})$, $\mathbf{u}_e(\boldsymbol{\rho})=\mathbf{u}_e(\boldsymbol{\rho}+\mathbf{R})$, where $\mathbf{R}$ stands for any lattice vectors.

In 2D PCs, we only need to investigate the in-plane propagating modes, i.e. $k_z=0$. Hence, the mode can be rewritten as

\begin{equation} \label{EQ6}
\begin{array}{c}
\mathbf{E}(\boldsymbol{\rho})=e^{-i \mathbf{k}_{ \|} \cdot \boldsymbol{\rho}}  \mathbf{u}_{e}(\boldsymbol{\rho}),\\
\mathbf{H}(\boldsymbol{\rho})=e^{-i \mathbf{k}_{ \|} \cdot \boldsymbol{\rho}}  \mathbf{u}_{h}(\bm{\rho}).
\end{array}
\end{equation}

Then, the Bloch's theorem can be written as

\begin{equation} \label{EQ7}
\begin{array}{c}
\mathbf{E}(\boldsymbol{\rho}+\mathbf{R})=e^{-i \mathbf{k}_{ \|} \cdot \mathbf{R}} \mathbf{E}(\boldsymbol{\rho}),\\
\mathbf{H}(\boldsymbol{\rho}+\mathbf{R})=e^{-i \mathbf{k}_{ \|} \cdot \mathbf{R} } \mathbf{H}(\boldsymbol{\rho}).
\end{array}
\end{equation}

\subsection{Discretization of the eigenvalue problems}

From the Maxwell's curl equations (equation \eqref{EQ1}), we can derive the governing equations for the modes of the PCs:

\begin{equation} \label{EQ8}
\nabla\times\hat{\mathbf{\mu}}^{-1}({\bf r})\nabla\times {\bf E}({\bf r}) = \omega^2 \hat{\epsilon}({\bf r}) {\bf E}({\bf r}),
\end{equation}
\begin{equation} \label{EQ9}
\nabla\times\hat{\mathbf{\epsilon}}^{-1}({\bf r})\nabla\times {\bf H}({\bf r}) = \omega^2 \hat{\mu}({\bf r}) {\bf H}({\bf r}).
\end{equation}

In 2D PCs, the modes can be separated into two distinct polarizations, transverse-magnetic (TM) modes with $z$-polarized electric fields and transverse-electric (TE) modes with $z$-polarized magnetic fields. The generalized eigenvalue matrix equations for the TM and TE modes of the 2D PCs can be discretized as,

\begin{equation} \label{EQ12}
\left\{U_{1}\left(\mu_{21}^{-1} V_{2}-\mu_{22}^{-1} V_{1}\right)-U_{2}\left(\mu_{11}^{-1} V_{2}-\mu_{12}^{-1} V_{1}\right)\right\} \hat{E}_{z}=\varepsilon_{33} k_{0}^{2} \hat{E}_{z}
\end{equation}
\begin{equation} \label{EQ13}
\left\{V_{1}\left(\varepsilon_{21}^{-1} U_{2}-\varepsilon_{22}^{-1} U_{1}\right)-V_{2}\left(\varepsilon_{11}^{-1} U_{2}-\varepsilon_{12}^{-1} U_{1}\right)\right\} \hat{H}_{z}=\mu_{33} k_{0}^{2} \hat{H}_{z}
\end{equation}

\begin{equation} \label{EQ14}
{{U}_{1}}=\frac{1}{{{Q}_{1}}}\left[ \begin{matrix}
   -1 & 1 & {} & {} & {} & {} & {} & {} & {}  \\
   {} & \ddots  & \ddots  & {} & {} & {} & {} & {} & {}  \\
   {{u}_{x}} & {} & -1 & 0 & {} & {} & {} & {} & {}  \\
   {} & {} & {} & -1 & 1 & {} & {} & {} & {}  \\
   {} & {} & {} & {} & \ddots  & \ddots  & {} & {} & {}  \\
   {} & {} & {} & {{u}_{x}} & {} & -1 & 0 & {} & {}  \\
   {} & {} & {} & {} & {} & {} & \ddots  & 1 & {}  \\
   {} & {} & {} & {} & {} & {} & {} & -1 & \ddots   \\
   {} & {} & {} & {} & {} & {} & {{u}_{x}} & {} & -1  \\
\end{matrix} \right]
\end{equation}

\begin{equation} \label{EQ14b}
{{V}_{1}}=\frac{1}{{{Q}_{1}}}\left[ \begin{matrix}
   1 & {} & {{v}_{x}} & {} & {} & {} & {} & {} & {}  \\
   -1 & 1 & {} & {} & {} & {} & {} & {} & {}  \\
   {} & \ddots  & \ddots  & {} & {} & {} & {} & {} & {}  \\
   {} & {} & 0 & 1 & {} & {{v}_{x}} & {} & {} & {}  \\
   {} & {} & {} & -1 & 1 & {} & {} & {} & {}  \\
   {} & {} & {} & {} & \ddots  & \ddots  & {} & {} & {}  \\
   {} & {} & {} & {} & {} & 0 & 1 & {} & {{v}_{x}}  \\
   {} & {} & {} & {} & {} & {} & -1 & \ddots  & {}  \\
   {} & {} & {} & {} & {} & {} & {} & \ddots  & 1  \\
\end{matrix} \right]
\end{equation}

\begin{equation} \label{EQ14c}
{{U}_{2}}=\frac{1}{{{Q}_{2}}}\left[ \begin{matrix}
   -1 & {} & {} & 1 & {} & {} & {} & {} & {}  \\
   {} & -1 & {} & {} & 1 & {} & {} & {} & {}  \\
   {} & {} & \ddots  & {} & {} & \ddots  & {} & {} & {}  \\
   {} & {} & {} & -1 & {} & {} & \ddots  & {} & {}  \\
   {} & {} & {} & {} & \ddots  & {} & {} & 1 & {}  \\
   {} & {} & {} & {} & {} & -1 & {} & {} & 1  \\
   {{u}_{y}} & {} & {} & {} & {} & {} & -1 & {} & {}  \\
   {} & \ddots  & {} & {} & {} & {} & {} & \ddots  & {}  \\
   {} & {} & {{u}_{y}} & {} & {} & {} & {} & {} & -1  \\
\end{matrix} \right]
\end{equation}

\begin{equation} \label{EQ14d}
{{V}_{2}}=\frac{1}{{{Q}_{2}}}\left[ \begin{matrix}
   1 & {} & {} & {} & {} & {} & {{v}_{y}} & {} & {}  \\
   {} & \ddots  & {} & {} & {} & {} & {} & {{v}_{y}} & {}  \\
   {} & {} & 1 & {} & {} & {} & {} & {} & {{v}_{y}}  \\
   -1 & {} & {} & 1 & {} & {} & {} & {} & {}  \\
   {} & {} & {} & {} & \ddots  & {} & {} & {} & {}  \\
   {} & {} & \ddots  & {} & {} & {} & {} & {} & {}  \\
   {} & {} & {} & \ddots  & {} & {} & \ddots  & {} & {}  \\
   {} & {} & {} & {} & {} & {} & {} & {} & {}  \\
   {} & {} & {} & {} & {} & -1 & {} & {} & 1  \\
\end{matrix} \right]
\end{equation}

The grid points at the boundaries are treated by the Bloch's theorem, consequently

\begin{equation}\label{EQ15a}
u_{x}=\exp \left(i \mathbf{k} \cdot a_{1} \hat{\bm{u}}_{1}\right), v_{x}=-\exp \left(-i \mathbf{k} \cdot a_{1} \hat{\bm{u}}_{1}\right), \\
u_{y}=\exp \left(i \mathbf{k} \cdot a_{2} \hat{\bm{u}}_{2}\right), v_{y}=-\exp \left(-i \mathbf{k} \cdot a_{2} \hat{\bm{u}}_{2}\right),
\end{equation}
where $a_1$, $a_2$ are the lengths of the unit cell along the directions of $\hat{\bm{u}}_{1}$ and $\hat{\bm{u}}_{2}$.

By solving these eigenvalue equations, the band structure together with the eigenstates of 2D PCs can be easily obtained.
The matrices
\begin{equation}\label{EQ15b}
\Theta_\mathbf{E}=\left\{U_{1}\left(\mu_{21}^{-1} V_{2}-\mu_{22}^{-1} V_{1}\right)-U_{2}\left(\mu_{11}^{-1} V_{2}-\mu_{12}^{-1} V_{1}\right)\right\}
\end{equation}
\begin{equation}\label{EQ15c}
\Theta_\mathbf{H}=\left\{V_{1}\left(\varepsilon_{21}^{-1} U_{2}-\varepsilon_{22}^{-1} U_{1}\right)-V_{2}\left(\varepsilon_{11}^{-1} U_{2}-\varepsilon_{12}^{-1} U_{1}\right)\right\}
\end{equation}
are discretized from the operator $\nabla\times\hat{\mathbf{\mu}}^{-1}({\bf r})\nabla\times$ and  $\nabla\times\hat{\mathbf{\epsilon}}^{-1}({\bf r})\nabla\times$. The condition of the electromagnetic modes $\mathbf{E}_{\omega_1}, \mathbf{E}_{\omega_2}$(or $\mathbf{H}_{\omega_1}, \mathbf{H}_{\omega_2}$) orthogonality is the Hermitian inner product satisfies $\left\langle \mathbf{E}_{\omega_1} | \mathbf{E}_{\omega_2} \right\rangle=\int \epsilon(\mathbf{r}) \mathbf{E}_{\omega_1} ^{*} \cdot \mathbf{E}_{\omega_2}  d^{2} \mathbf{r}=0$ or $\left\langle \mathbf{H}_{\omega_1} | \mathbf{H}_{\omega_2} \right\rangle=\int \mu(\mathbf{r}) \mathbf{H}_{\omega_1} ^{*} \cdot \mathbf{H}_{\omega_2}  d^{2} \mathbf{r}=0$. The subscript ${\omega_1}$ and ${\omega_1}$ stand for the different eigen-frequency under a fixed wave-vector $\mathbf{k}$.

\section{Chern number calculation}

\subsection{Single-band Chern number}

Chern number is an integer which determines the topological classification of different materials or structures. Taking the TM mode of a 2D PC as an example, the Chern number of the $\mathrm{n}^{th}$ band can be computed by integrating the Berry curvature over the first Brillouin zone as following\cite{Fukui2005chern},

\begin{equation} \label{EQ16}
\begin{aligned}[b]
C^{(n)} &= \frac{1}{2 \pi} \int_{\mathrm{BZ}} \mathbf{F}_{n}(\mathbf{k}) \mathrm{d} \mathbf{k}\\
        &= \frac{1}{2 \pi} \int_{\mathrm{BZ}} \nabla_\mathbf{k}\times\mathbf{A}_{n}(\mathbf{k}) \mathrm{d} \mathbf{k}\\
        &=\frac{1}{2 \pi i} \oint_{\partial \mathrm{BZ}}\left\langle\mathbf{u}_{n,e,\mathbf{k}}\left|\nabla_{\mathbf{k}}\right| \mathbf{u}_{n,e,\mathbf{k}}\right\rangle \mathrm{d} \mathbf{k}.\\
\end{aligned}
\end{equation}
Here, the $\mathbf{F}_{n}(\mathbf{k})$ is the Berry curvature. $\mathbf{A}_{n}(\mathbf{k})=-i\left\langle\mathbf{u}_{n,e, \mathbf{k}}\left|\nabla_{\mathbf{k}}\right| \mathbf{u}_{n,e,\mathbf{k}}\right\rangle$ is the Berry connection.  $\mathbf{u}_{n,e,\mathbf{k}}$ denotes the normalized eigenstate satisfying $\left. \left|\mathbf{u}_{n,e,\mathbf{k}}( \boldsymbol{\rho}) \right. \right\rangle=\left. \left|e^{i\mathbf{k} \cdot \boldsymbol{\rho}} \mathbf{E}_{n,e,\mathbf{k}}( \boldsymbol{\rho})\right.   \right\rangle $  and  $\left. \left|\mathbf{E}_{n,e,\mathbf{k}}( \boldsymbol{\rho})\right.   \right\rangle $ is the eigenstate solved from equation \eqref{EQ12}.

\begin{figure}[htbp]
	\centering
	\includegraphics[width=11cm]{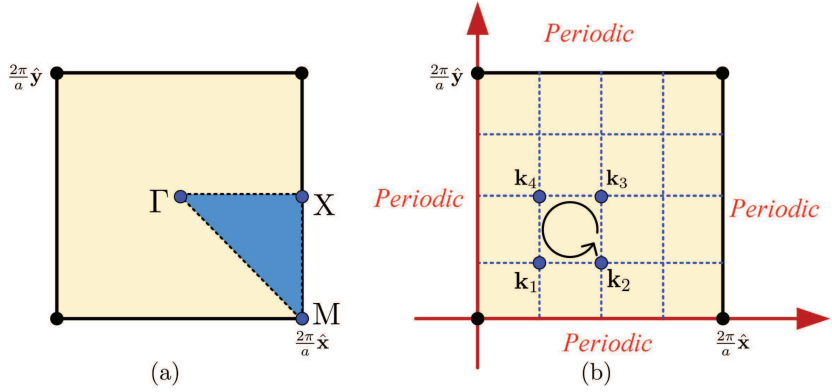}
	\caption{Computational scheme for the calculation of Chern numbers. (a) A square-shaped first Brillouin zone; (b) Discretization of the first Brillouin zone, and the computational scheme for the Chern number. }
	\label{FIG2}
\end{figure}

For numerical discretization, the Chern number is calculated in the discretized Brillouin zone.
In Fig.~\ref{FIG2}(a), an example for a square-shaped first Brillouin zone is shown. It is discretized into $4\times4$ plaquettes.
Then, the Chern number is calculated using

\begin{equation} \label{EQ17}
\begin{aligned}[b]
C^{(n)} &= \frac{1}{2 \pi} \int_{\mathrm{BZ}} \mathbf{F}_{n}(\mathbf{k}) \mathrm{d} \mathbf{k}\\
        &=\frac{1}{2 \pi \mathrm{i}} \oint_{\partial \mathrm{BZ}}\left\langle\mathbf{u}_{n,e,\mathbf{k}}\left|\nabla_{\mathbf{k}}\right| \mathbf{u}_{n,e,\mathbf{k}}\right\rangle \mathrm{d} \mathbf{k}\\
        &= \frac{1}{2 \pi} \sum_{\mathbf{k}\in\mathrm{BZ}} F_{\mathbf{k}}^{(n)} \Delta S_{\mathbf{k}} \\
        &{= \frac{1}{2 \pi} \sum_{\mathbf{k}\in\mathrm{BZ}} \operatorname{Im} \ln\left[{
        U_{\mathbf{k}_1 \rightarrow \mathbf{k}_2}^{(n)} U_{\mathbf{k}_2 \rightarrow \mathbf{k}_3 }^{(n)}  U_{\mathbf{k}_3\rightarrow \mathbf{k}_4}^{(n)}
        U_{\mathbf{k}_4 \rightarrow \mathbf{k}_1}^{(n)}
        }\right]}
\end{aligned}
\end{equation}
where $U_{\mathbf{k}_{\alpha} \rightarrow \mathbf{k}_{\beta}}^{(n)} \equiv \frac{\left\langle\mathbf{u}_{n,e,\mathbf{k}_{\alpha}} | \mathbf{u}_{n,e, \mathbf{k}_{\beta}}\right\rangle}{\left|\left\langle\mathbf{u}_{n,e,\mathbf{k}_{\alpha}} | \mathbf{u}_{n,e, \mathbf{k}_{\beta}}\right\rangle\right|}$, $\alpha,\beta=1,2,3,4$, and $\left\langle\mathbf{u}_{n,e,\mathbf{k}_{\alpha}} | \mathbf{u}_{n,e, \mathbf{k}_{\beta}}\right\rangle=\int \epsilon(\mathbf{r}) \mathbf{u}_{n,e,\mathbf{k}_{\alpha}}(\mathbf{r}) ^{*} \cdot \mathbf{u}_{n,e, \mathbf{k}_{\beta}}(\mathbf{r})  d^{2} \mathbf{r}.$
$\mathbf{k}_1,\mathbf{k}_2,\mathbf{k}_3,\mathbf{k}_4$ are the vertices of the each plaquette. The  integration is replaced by a summation $\sum_{\mathbf{k}\in\mathrm{BZ}} F_{\mathbf{k}}^{(n)} \Delta S_{\mathbf{k}}$.

For the first Brillouin zone in different shape, similar discretization scheme can be used. It should be noted that to simplify the numerical calculation, we may need to use the equivalent Brillouin zone which shares the same size with the first Brillouin zone but in a parallelogram shape with its edges coinciding with the reciprocal lattice vector \cite{Jin2017Infrared}.

\subsection{Composite Chern number}

When the bands are degenerate, the Chern number can not be assigned to each band, so equation \eqref{EQ17} will no longer be valid. For this case, these degenerate bands (taking the $\{n, n+1, \ldots, n+N-1\}$ bands as an example) jointly share a composite (first) Chern number, $C^{(n \oplus n+1 \oplus \cdots \oplus n+N-1)}$ which is associated with the multiplets $\mathbf{u}_{n \mathbf{k}} \oplus \mathbf{u}_{n+1, \mathbf{k}} \oplus \cdots \oplus \mathbf{u}_{n+N-1, \mathbf{k}}$. It is calculated using \cite{Fukui2005chern,Wang2015Topological}

\begin{equation} \label{EQ19}
\begin{aligned}[b]
C^{(n \oplus n+1 \oplus \cdots \oplus n+N-1)} &= \frac{1}{2 \pi} \operatorname{Tr} \oint_{\partial \mathrm{BZ}} \mathbf{A}_{\mathbf{k}}^{(n \oplus n+1 \oplus \cdots \oplus n+N-1)} \mathrm{d} \mathbf{k} \\
&=\frac{1}{2 \pi \mathrm{i}} \oint_{\partial \mathrm{BZ}}
\left[ \begin{matrix}
  {\ddots} & {\vdots} & {\adots }  \\
  {\cdots} & {\left\langle\mathbf{u}_{n_l,e,\mathbf{k}}\left|\nabla_{\mathbf{k}}\right| \mathbf{u}_{n_m,e,\mathbf{k}}\right\rangle } & {\cdots}  \\
  {\adots } & {\vdots} & {\ddots}\\
\end{matrix} \right]
 \mathrm{d} \mathbf{k}
\end{aligned}
\end{equation}
Here, $\mathbf{A}_{\mathbf{k}}^{(n \oplus n+1 \oplus \cdots \oplus n+N-1)}$ is an $N\times N$ matrix with the $(l,m)$ element, $-\mathrm{i}{\left\langle\mathbf{u}_{n_l,e,\mathbf{k}}\left|\nabla_{\mathbf{k}}\right| \mathbf{u}_{n_m,e,\mathbf{k}}\right\rangle} $, and $n_{l}=n+l-1, n_{m}=n+m-1$.

By applying Stokes' theorem, the composite Chern number can be rewritten in terms of the composite Berry flux:

\begin{equation} \label{EQ20}
\begin{aligned}[b]
C^{(n \oplus n+1 \oplus \cdots \oplus n+N-1)}&=\frac{1}{2 \pi} \int_{\mathrm{BZ}} F_{\mathbf{k}}^{(n \oplus n+1 \oplus \cdots \oplus n+N-1)} \mathrm{d} S_{\mathbf{k}} \\
&=\frac{1}{2 \pi} \int_{\mathrm{BZ}} \operatorname{Tr} \nabla_{\mathbf{k}} \times \mathbf{A}_{\mathbf{k}}^{(n \oplus n+1 \oplus \cdots \oplus n+N-1)} \mathrm{d} S_{\mathbf{k}},
\end{aligned}
\end{equation}

\begin{equation} \label{EQ21}
\begin{aligned}[b]
&F_{\mathbf{k}}^{(n \oplus n+1 \oplus \cdots \oplus n+N-1)} \mathrm{d} S_{\mathbf{k}}\\
&=\operatorname{Im} \mathrm{Tr}
\left\{
\ln \left[  \mathbf{U}_{\mathbf{k}_1 \rightarrow \mathbf{k}_2}^{(n \oplus \cdots \oplus n+N-1)}
  \mathbf{U}_{\mathbf{k}_2 \rightarrow \mathbf{k}_3}^{(n \oplus \cdots \oplus n+N-1)}
  \mathbf{U}_{\mathbf{k}_3 \rightarrow \mathbf{k}_4}^{(n \oplus \cdots \oplus n+N-1)}
  \mathbf{U}_{\mathbf{k}_4 \rightarrow \mathbf{k}_1}^{(n \oplus \cdots \oplus n+N-1)}
 \right]
\right\}\\
&=\operatorname{Im} \ln
\left\{
\det \left[  \mathbf{U}_{\mathbf{k}_1 \rightarrow \mathbf{k}_2}^{(n \oplus \cdots \oplus n+N-1)}
  \mathbf{U}_{\mathbf{k}_2 \rightarrow \mathbf{k}_3}^{(n \oplus \cdots \oplus n+N-1)}
  \mathbf{U}_{\mathbf{k}_3 \rightarrow \mathbf{k}_4}^{(n \oplus \cdots \oplus n+N-1)}
  \mathbf{U}_{\mathbf{k}_4 \rightarrow \mathbf{k}_1}^{(n \oplus \cdots \oplus n+N-1)}
 \right]
\right\}.
\end{aligned}
\end{equation}

The element $(l,m)$ of the link matrix $\mathbf{U}_{\mathbf{k}_{\alpha} \rightarrow \mathbf{k}_{\beta}}^{(n \oplus \cdots \oplus n+N-1)}$ is

\begin{equation} \label{EQ22}
\mathbf{U}_{\mathbf{k}_{\alpha} \rightarrow \mathbf{k}_{\beta}}^{(n \oplus \cdots \oplus n+N-1)}\left(l,m\right)=
\frac{\left\langle\mathbf{u}_{n_l,e,\mathbf{k}_{\alpha}} | \mathbf{u}_{n_m,e, \mathbf{k}_{\beta}}\right\rangle}{\left|\left\langle\mathbf{u}_{n_l,e,\mathbf{k}_{\alpha}} | \mathbf{u}_{n_m,e, \mathbf{k}_{\beta}}\right\rangle\right|}.
\end{equation}

Different from the single-band Chern number, in the composite Chern number, the link matrix $\mathbf{U}_{\mathbf{k}_{\alpha} \rightarrow \mathbf{k}_{\beta}}^{(n \oplus \cdots \oplus n+N-1)}$ is used to replace the link variable $U_{\mathbf{k}_{\alpha} \rightarrow \mathbf{k}_{\beta}}^{(n)}$. In the discretized Brillouin zone, the Chern number is expressed as

\begin{equation} \label{EQ23}
\begin{aligned}[b]
&C^{(n \oplus n+1 \oplus \cdots \oplus n+N-1)} \\
&= \frac{1}{2 \pi}\sum_{\mathrm{BZ}}
\operatorname{Im} \ln
\left\{
\det \left[  \mathbf{U}_{\mathbf{k}_1 \rightarrow \mathbf{k}_2}^{(n \oplus \cdots \oplus n+N-1)}
  \mathbf{U}_{\mathbf{k}_2 \rightarrow \mathbf{k}_3}^{(n \oplus \cdots \oplus n+N-1)}
  \mathbf{U}_{\mathbf{k}_3 \rightarrow \mathbf{k}_4}^{(n \oplus \cdots \oplus n+N-1)}
  \mathbf{U}_{\mathbf{k}_4 \rightarrow \mathbf{k}_1}^{(n \oplus \cdots \oplus n+N-1)}
 \right]
\right\}
\end{aligned}
\end{equation}

As for the calculation of the Chern number of TE modes, the derived formulae are still valid, except that the normalized eigenstates now become the normalized magnetic field, $\mathbf{u}_{n,h,\mathbf{k}_{\alpha}}(\mathbf{r})$. Finally, we would like to note that the algorithm for the Chern number calculations described above guarantees the calculated Chern number is strictly an integer for arbitrary discretization spacing (for the detailed proof, see~\cite{Fukui2005chern}). However, this does not mean that any discretization will give the correct Chern numbers, but rather, convergence to the correct Chern numbers still requires a sufficient sampling of the Brillouin zone, which can be achieved using rather coarse discretization as we will show in the following section.

\section{Numerical results}

In the first numerical example, a 2D PC composed of Yttrium-Iron-Garnet (YIG) rods ($\epsilon=15\epsilon_0$) in a square lattice is analyzed. The radius of the rods is $0.11a$, where $a$ is the lattice constant. An external direct current (DC) magnetic field is applied in the $z$ direction (out-of-plane) to break the time-reversal symmetry. The permeability is a tensor matrix written as

\begin{equation} \label{EQ24}
\bar{\bar{\mu}}=\left[ \begin{array}{ccc}{\mu} & {i \kappa} & {0} \\ {-i \kappa} & {\mu} & {0} \\ {0} & {0} & {\mu_{0}}\end{array}\right],\quad
\mu=1+\frac{\omega_m\omega_0}{\omega^2_0-\omega^2},\quad \kappa=\frac{\omega_m\omega}{\omega^2_0-\omega^2},
\end{equation}
where $\omega_0=\gamma H_0$ is the precession frequency and $\omega_m=4\pi{\gamma}{M_s}$.
When the external magnetic field is 1600 Gauss at 4.28 GHz, $\kappa=12.4 \mu_{0}$, and $\mu=14 \mu_{0}$ \cite{wang2008reflection,Pozar}.

\begin{figure}[htbp]
\centering
\subfigure[The band structure of the TM modes.]{
\begin{minipage}[t]{0.45\linewidth}
\centering
\includegraphics[width=5.5cm]{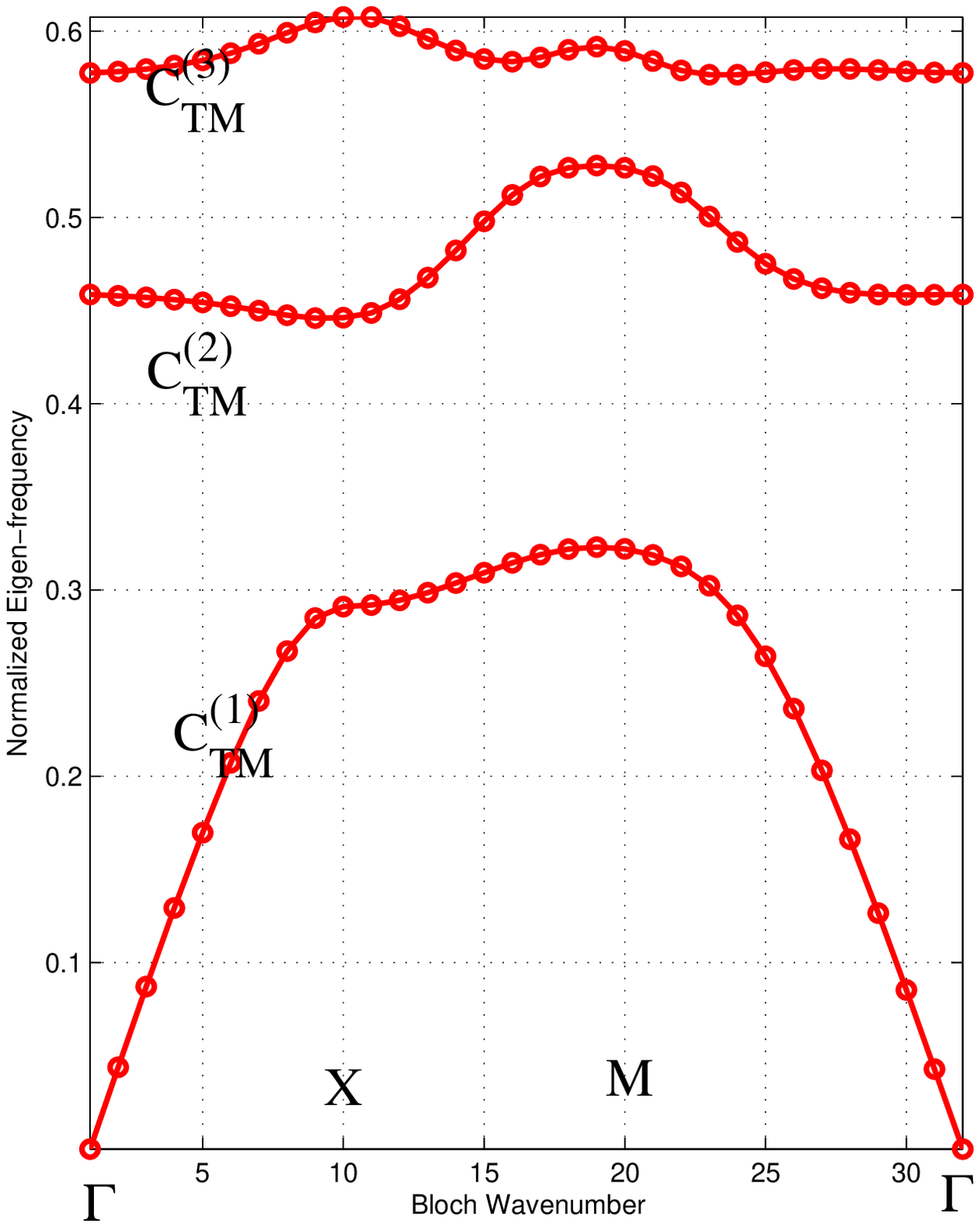}
\end{minipage}%
}%
\subfigure[The band structure of the TE modes.]{
\begin{minipage}[t]{0.45\linewidth}
\centering
\includegraphics[width=5.5cm]{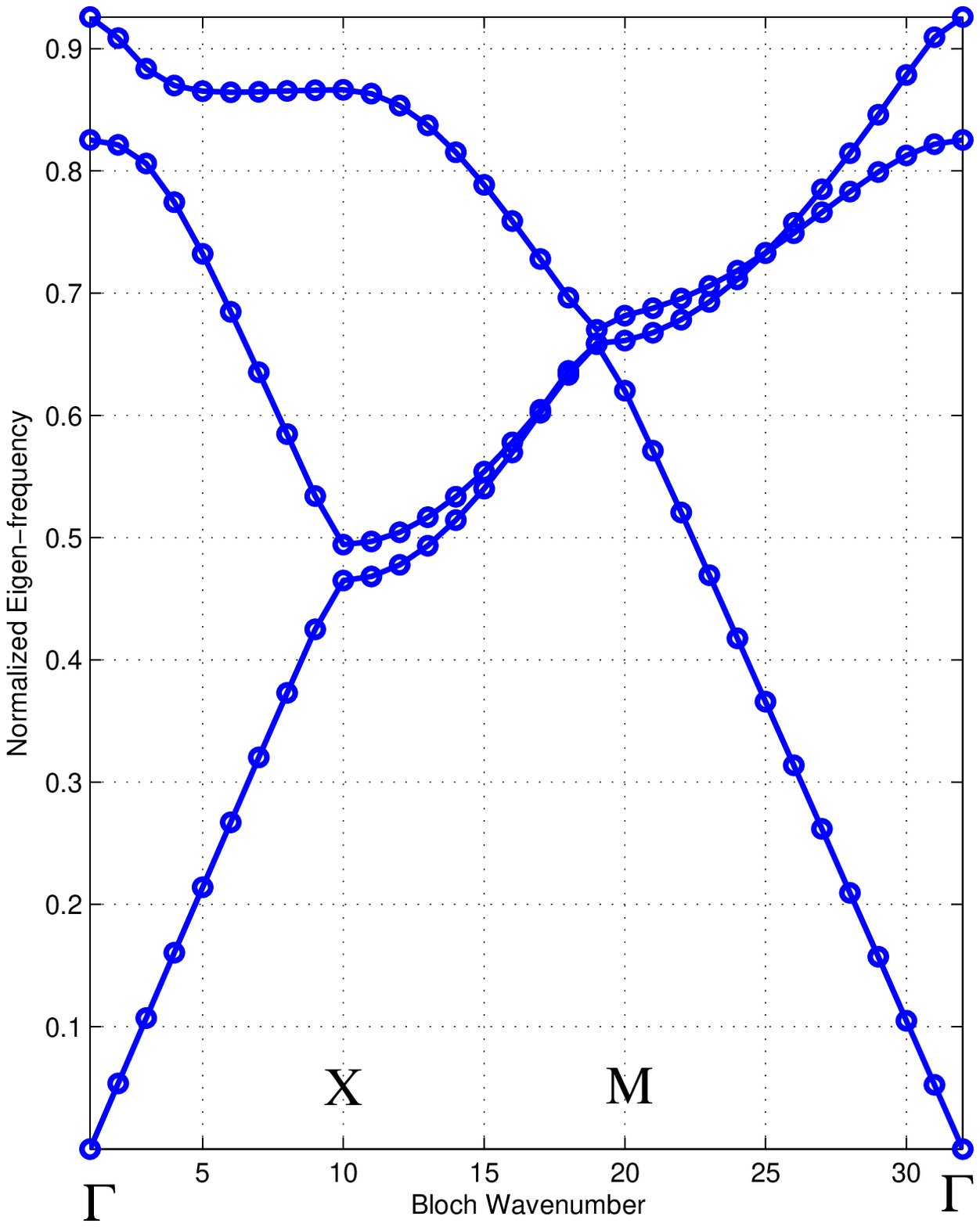}
\end{minipage}%
}%
\centering
\caption{The Band structures of a 2D PC composed of YIG rods in a square lattice when a 1600 Gauss $+\mathrm{z}$ DC magnetic field is applied. $C^{(1)}_\mathrm{TM}$, $C^{(2)}_\mathrm{TM}$, $C^{(3)}_\mathrm{TM}$  denote the Chern numbers of the TM modes of the associated bands.}
\label{FIG3}
\end{figure}

Firstly, we calculate the band structure of the PC using the FDFD method. In order to ensure the accuracy, the mesh size is set to be $a/100$. The calculated band structures for the TM modes and TE modes are plotted in Figs. \ref{FIG3}(a) and Figs. \ref{FIG3}(b), respectively. In Fig. \ref{FIG3}(a), $C^{(1)}_\mathrm{TM}$, $C^{(2)}_\mathrm{TM}$, $C^{(3)}_\mathrm{TM}$ stand for the Chern numbers of the first, second and third bands.  Then, we investigate the computational accuracy of the Chern number by discretizing the Brillouin zone with different numbers of plaquettes. The computational results are shown in Table \ref{tab:table1}. We can see that the computational results of Chern numbers are converged when the number of plaquettes is larger than $4 \times 4$. By further checking the values of the Chern numbers and composite Chern numbers, we find $C^{(1\oplus2)}_\mathrm{TM}=C^{(1)}_\mathrm{TM}+C^{(2)}_\mathrm{TM}$,
and $C^{(1\oplus2\oplus3)}_\mathrm{TM}=C^{(1)}_\mathrm{TM}+C^{(2)}_\mathrm{TM}+C^{(3)}_\mathrm{TM}$, which is as expected.


\begin{table}[ht]
\centering
\caption{\label{tab:table1}The TM-mode Chern numbers and composite Chern numbers of different bands calculated with different mesh sizes of the first Brillouin zone.}
\begin{tabular}{ c c c c c c  }
\hline
\hline
\diagbox{Mesh }{Chern}& $C^{(1)}_\mathrm{TM}$ & $C^{(2)}_\mathrm{TM}$ & $C^{(3)}_\mathrm{TM}$  & $C^{(1\oplus2)}_\mathrm{TM}$ & $C^{(1\oplus2\oplus3)}_\mathrm{TM}$  \\
\hline
$2\times2$& 0 & 1 & 0  & 1 & -1  \\
\hline
$4\times4$& 0 & 1 & -2 & 1 & -1  \\
\hline
$8\times8$& 0 & 1 & -2 & 1 & -1  \\
\hline
$16\times16$& 0 & 1 & -2 & 1 & -1 \\
\hline
\hline
\end{tabular}
\end{table}

In the second numerical example, a 2D PC composed of gyrotropic rods in a square lattice is analyzed. The geometric parameters are the same as these of the first example ($r=0.11a$). The relative permittivity and permeability are expressed as following,

\begin{equation} \label{EQ25}
\bar{\bar{\varepsilon}}=\left[ \begin{array}{ccc}{\varepsilon_{d}} & {i \varepsilon_{f}} & {0} \\ {-i \varepsilon_{f}} & {\varepsilon_{d}} & {0} \\ {0} & {0} & {\varepsilon_\perp}\end{array}\right], \bar{\bar{\mu}}=\left[ \begin{array}{ccc}{\mu_{d}} & { i\mu_{f}} & {0} \\ {-i \mu_{f}} & {\mu_{d}} & {0} \\ {0} & {0} & {\mu_{_\perp}}\end{array}\right].
\end{equation}
Here, the parameters are chosen to satisfy the relationship $\varepsilon_{d}=\mu_{d}, \varepsilon_{f}=-\mu_{f}, \varepsilon_{\perp}=\mu_{\perp}$. Referring to the YIG material under an external magnetic field at 4.28 GHz, the corresponding parameters are set as $\varepsilon_{d}=14, \varepsilon_{f}=-12.4$ and $\mu_{\perp}=15$, $\mu_{d}=14, \mu_{f}=12.4$ and $\varepsilon_{\perp}=15$ \cite{sun2019Photonic}.

\begin{figure}[htbp]
\centering
\subfigure[The band structure of the TM modes.]{
\begin{minipage}[t]{0.45\linewidth}
\centering
\includegraphics[width=5.0cm]{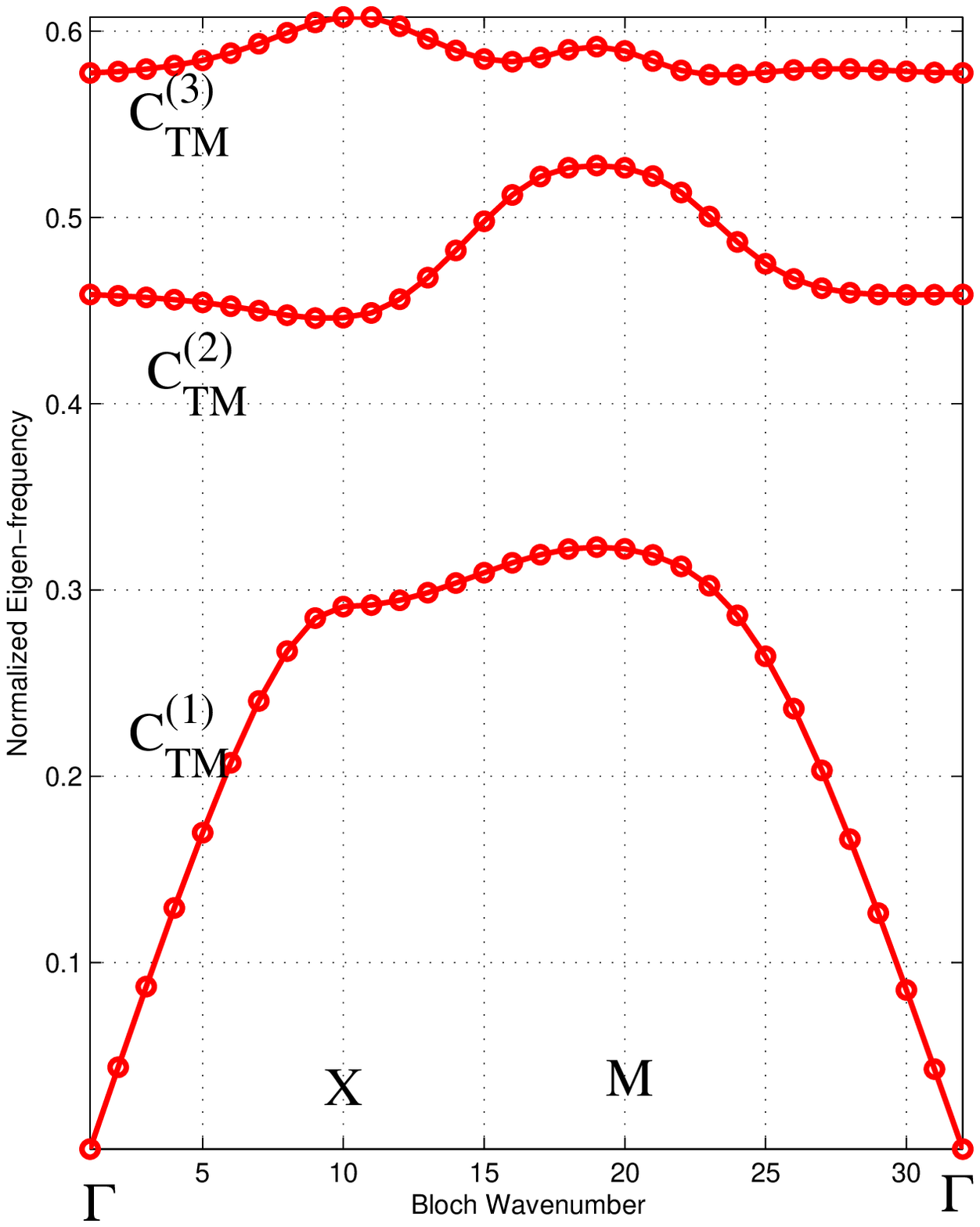}
\end{minipage}%
}%
\subfigure[The band structure of the TE modes.]{
\begin{minipage}[t]{0.45\linewidth}
\centering
\includegraphics[width=5.0cm]{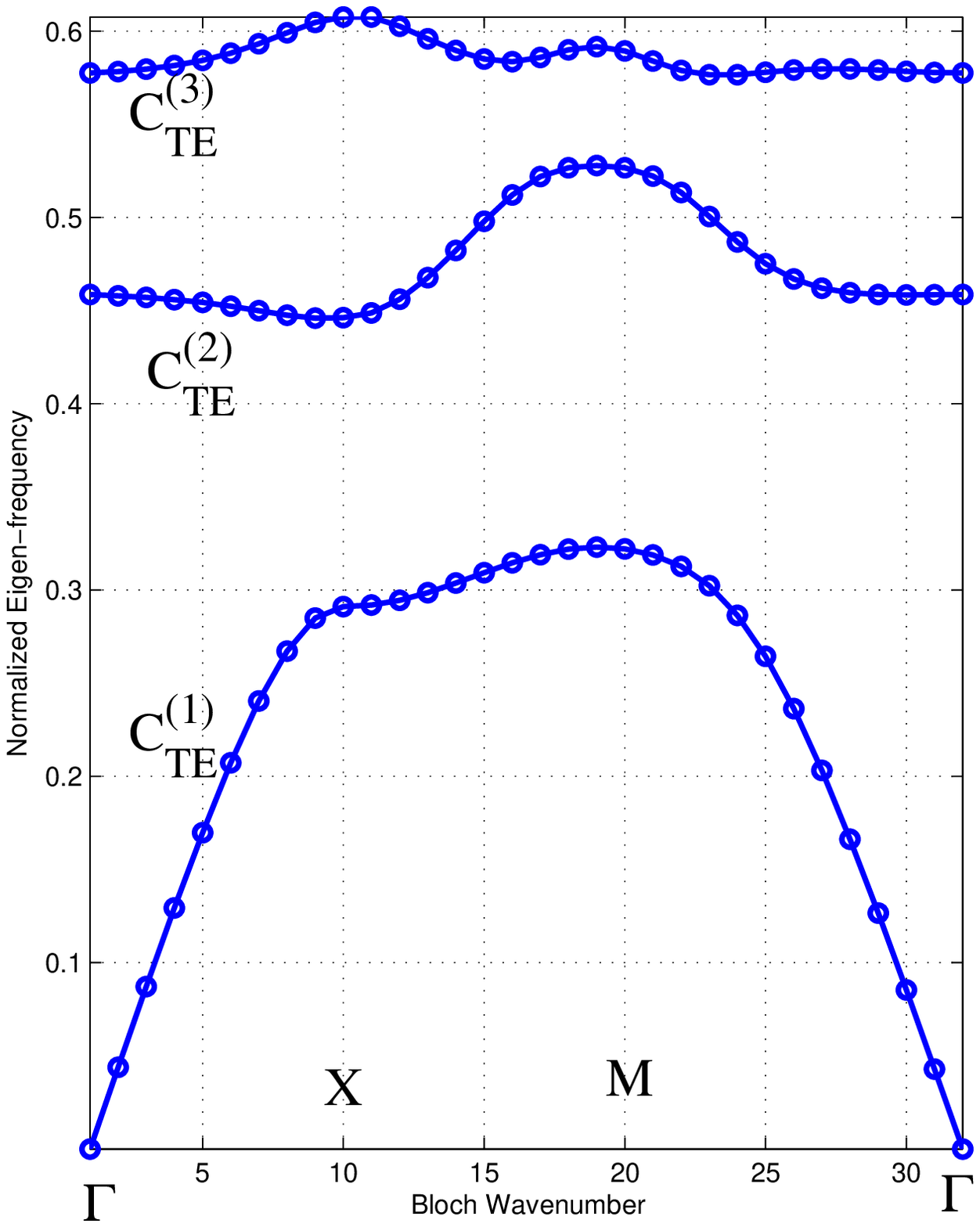}
\end{minipage}%
}%
\centering
\caption{The Band structures of a 2D PC composed of gyrotopic rods in a square lattice when a 1600 Gauss $+\mathrm{z}$ DC magnetic field is applied. $C^{(1)}_\mathrm{TM/TE}$, $C^{(2)}_\mathrm{TM/TE}$, $C^{(3)}_\mathrm{TM/TE}$  denote the Chern numbers of the TM/TE modes of the associated bands.}
\label{FIG4}
\end{figure}

The band structures of TM and TE modes are shown in Fig. \ref{FIG4}. Different from the first numerical example, by using the gyrotropic material, there are band gaps for both the TM and TE modes. The $C^{(1)}_\mathrm{TM/TE}$, $C^{(2)}_\mathrm{TM/TE}$, $C^{(3)}_\mathrm{TM/TE}$ stand for the Chern numbers of the first, second and third bands for the TM/TE mode.  As shown in the Table 2 and Table 3, $4\times4$ meshes in the Brillouin zone are also sufficient for the Chern number computation for TM/TE modes. With the increase of the discretization density, the calculated results are stable.

\begin{table}[htbp]
\centering
\caption{\label{tab:table2}The TM-mode Chern numbers and composite Chern numbers of different bands calculated with different mesh sizes of the first Brillouin zone.}
\begin{tabular}{ c c c c c c  }
\hline
\hline
\diagbox{Mesh }{Chern}& $C^{(1)}_\mathrm{TM}$ & $C^{(2)}_\mathrm{TM}$ & $C^{(3)}_\mathrm{TM}$
& $C^{(1\oplus2)}_\mathrm{TM}$ & $C^{(1\oplus2\oplus3)}_\mathrm{TM}$ \\
\hline
$2\times2$& 0 & 1 & 0  & 1 & -1 \\
\hline
$4\times4$& 0 & 1 & -2  & 1 & -1 \\
\hline
$8\times8$& 0 & 1 & -2 & 1 & -1 \\
\hline
$16\times16$& 0 & 1 & -2 & 1 & -1\\
\hline
\hline
\end{tabular}
\end{table}

\begin{table}[htbp]
\centering
\caption{\label{tab:table3}The TE-mode Chern numbers and composite Chern numbers of different bands calculated with different mesh sizes of the first Brillouin zone.}
\begin{tabular}{ c c c c c c  }
\hline
\hline
\diagbox{Mesh }{Chern}& $C^{(1)}_\mathrm{TE}$ & $C^{(2)}_\mathrm{TE}$ & $C^{(3)}_\mathrm{TE}$
& $C^{(1\oplus2)}_\mathrm{TE}$ & $C^{(1\oplus2\oplus3)}_\mathrm{TE}$ \\
\hline
$2\times2$& 0 & -1 & 0  & -1 & 1 \\
\hline
$4\times4$& 0 & -1 & 2  & -1 & 1 \\
\hline
$8\times8$& 0 & -1 & 2  & -1 & 1 \\
\hline
$16\times16$& 0 & -1 & 2 & -1 & 1 \\
\hline
\hline
\end{tabular}
\end{table}

\begin{figure}[htbp]
	\centering
	\includegraphics[width=8cm]{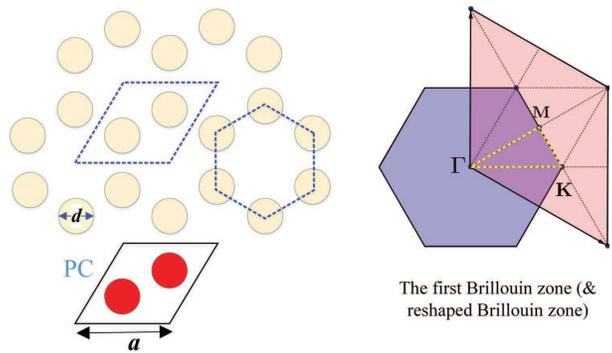}
	\caption{The unit cell and first Brillouin zone of the honeycomb lattice.}
	\label{FIG6}
\end{figure}

\begin{figure}[htbp]

	\centering
	\subfigure[Band structure of the TM modes for a 2D   honeycomb lattice.]{
		\begin{minipage}[t]{0.45\linewidth}
			\centering
			\includegraphics[width=5.0cm]{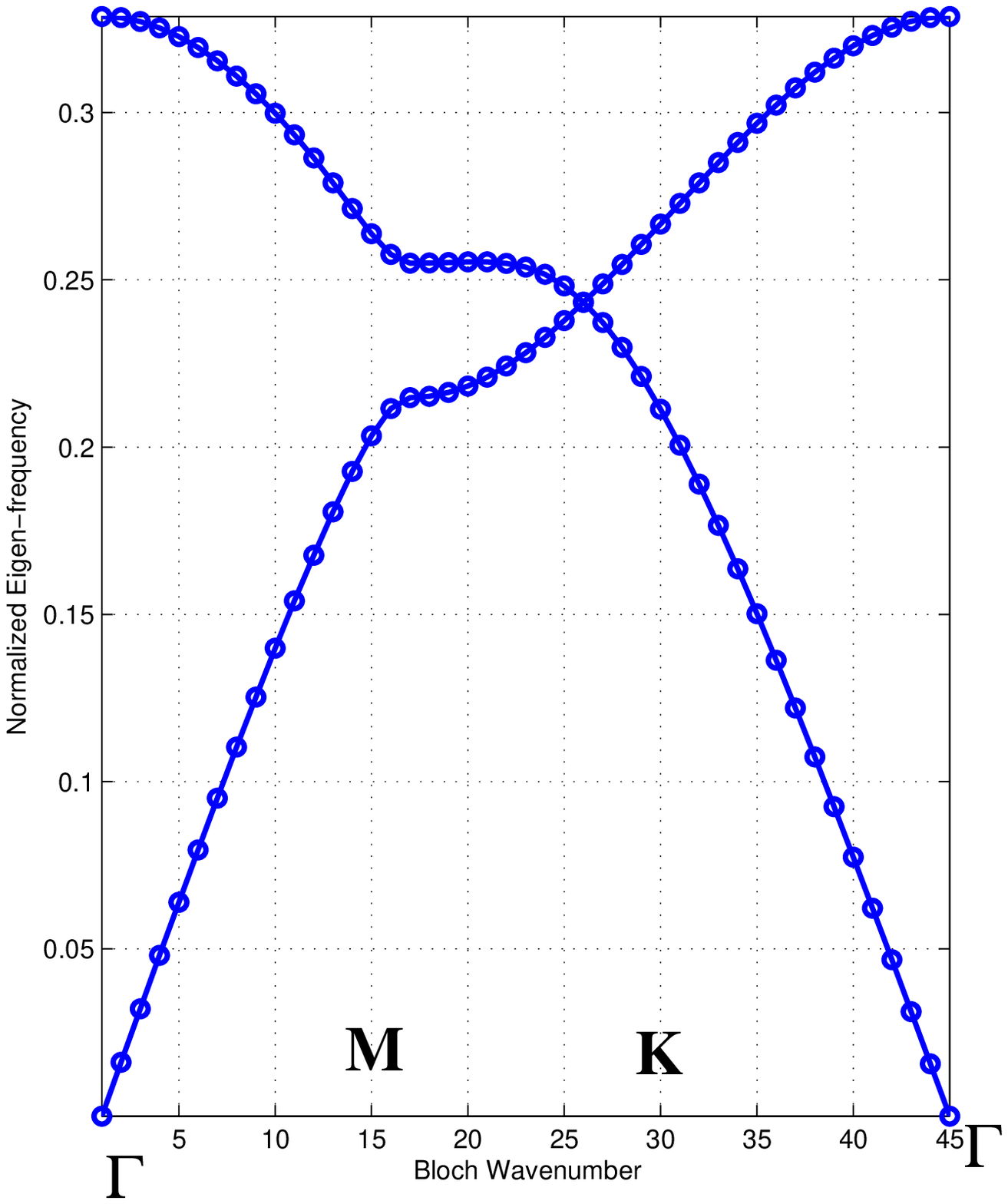}
		\end{minipage}%
	}%
	\subfigure[Band structure for honeycomb lattice of ferrite rods with $\varepsilon_r=15$, $ H_o=500$ $\mathrm{Oe}$, $4\pi M_s=1750$ $\mathrm{G}$, $r
=0.2a$, $a=10$ $\mathrm{mm}$, $f=7.5$ $\mathrm{GHz}$\cite{Ao2009One}. ]{
		\begin{minipage}[t]{0.45\linewidth}
			\centering
			\includegraphics[width=5.0cm]{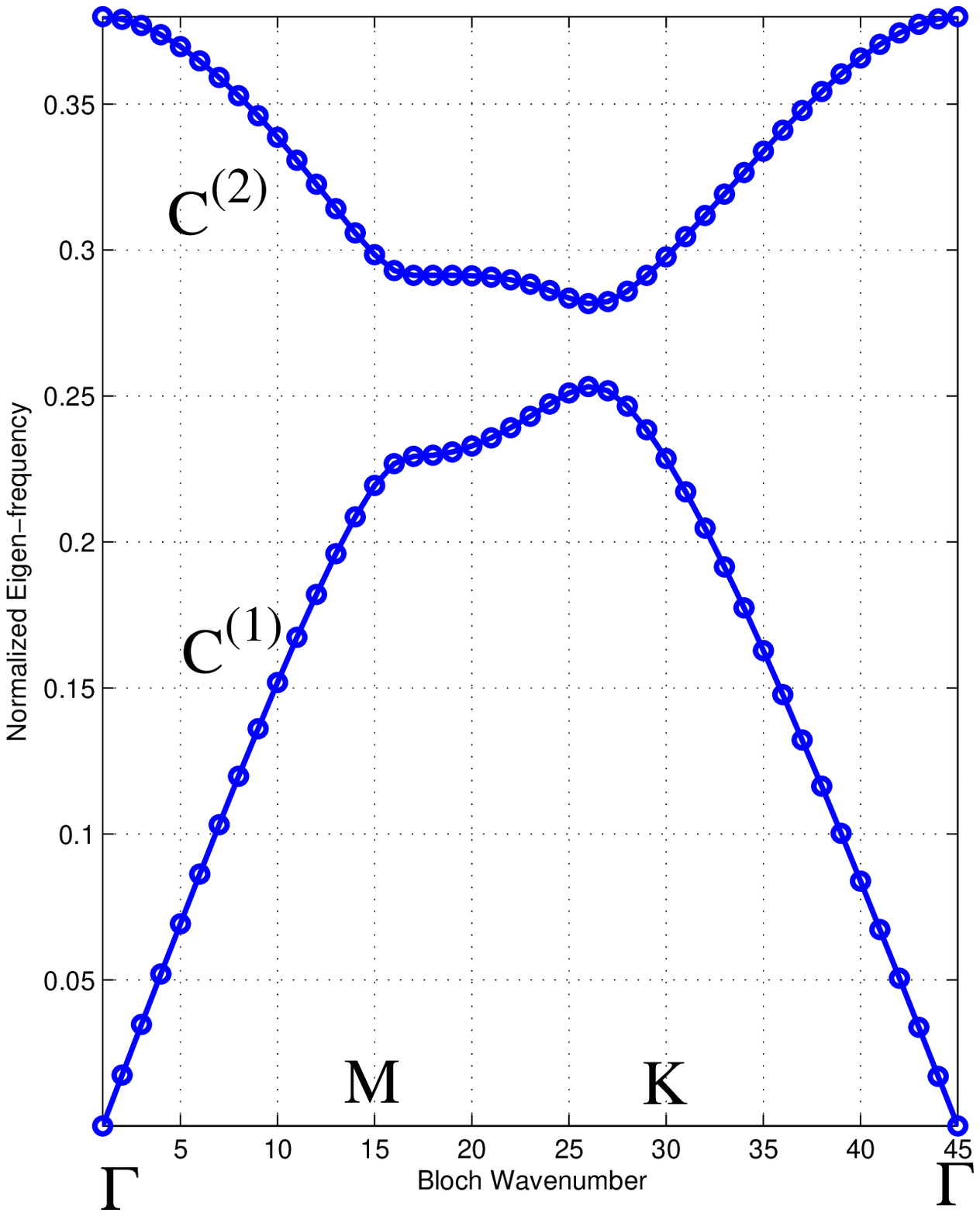}
		\end{minipage}%
	}%
	\centering
	\caption{ The band structure of 2D honeycomb lattice without and with DC magnetic field.}
	\label{FIG7}
\end{figure}

In the third numerical example, a 2D PC composed of YIG rods in a honeycomb lattice\cite{Ao2009One} is analyzed. The unit cell of the lattice is shown in the Fig. \ref{FIG6}.
The corresponding first Brillouin zone (hexagon region) constructed in the reciprocal lattice is also shown in the Fig. 5. As shown in this figure, for the convenience of discretization, the hexagon region needs to be equivalently transformed into a rhombus region, which can be discretized into smaller rhombus cells straightforwardly. The detailed reshaping and discretization procedure for the Brillouin zone can be found in ~\cite{Jin2017Infrared}.
In Fig. \ref{FIG7}(a), the first two TM bands with zero DC magnetic field are given. It can be seen that the two bands are degenerate at the $\mathbf{K}$ point. By applying a DC magnetic field, the time-reversal symmetry is broken and the degeneracy  is lifted, as shown in the Fig. \ref{FIG7}(b). The calculated Chern numbers and composite Chern numbers corresponding to the two bands are shown in the Table \ref{tab:table4}.  The non-zero Chern numbers can be used to explain the edge-state in \cite{Ao2009One} at 7.5 GHz. Also, as can be seen from this table, a very coarse mesh ($2\times2$) can guarantee the accuracy of the results.

\begin{table}[htbp]
	\centering
	\caption{\label{tab:table4}The honeycomb lattice's Chern numbers and composite Chern numbers of the first two bands calculated with different mesh sizes of the first Brillouin zone.}
	\begin{tabular}{ c c c c }
		\hline
		\hline
		\diagbox{Mesh }{Chern}& $C^{(1)}$ & $C^{(2)}$ & $C^{(1\oplus2)}$  \\
		\hline
		$2\times2$& 1 & -1 & 0 \\
		\hline
		$4\times4$&1 & -1 & 0\\
		\hline
		$8\times8$& 1 & -1 & 0\\
		\hline
		$16\times16$& 1 & -1 & 0\\
		\hline
		\hline
	\end{tabular}
\end{table}

\section{Conclusion}
In conclusion, a first-principle computational method for the Chern number of 2D gyrotropic PCs has been developed\cite{opensrc}. Firstly, the full-wave FDFD method was used to solve the generalized eigenvalue problem based on the Maxwell's equations. Then, the band structures of the PCs with different material properties were analyzed. The Chern numbers were calculated by integrating the Berry curvature over the first Brillouin zone. All the three numerical examples have demonstrated the accuracy and convergence of this method. Importantly, the method proposed in the current work could be potentially applied to 3D PCs~\cite{Oono15PRB} or 3D planar PCs when one cannot separate the TE and TM polarizations, in which case the full vectorial eigenmodes could be used for the Chern number calculations~\cite{haldane2008possible}.

\section*{Funding}
This work was supported in part by the National Natural Science Foundation of China (NSFC) (61722101, 61801002, 61901087, 61701424, 61975177); in part by Natural Science Foundation of Anhui Province (No.1808085QF183) and Key Natural Science Project of Anhui Provincial Education Department (KJ2018A0015).






\end{document}